\documentclass[aps,twocolumn,english,showpacs,floatfix]{revtex4}
\usepackage[T1]{fontenc}
\usepackage[latin1]{inputenc}
\usepackage{float}
\usepackage{graphicx}
\usepackage{amssymb}

\begin{document}

\title{Quantum Chaos and Regularity in Ultracold  Fermi Gases}

\author{M. Puig von Friesen, M. \"{O}gren and S. \AA berg}

\affiliation{Mathematical Physics, Lund University, P.O.Box 118,
SE-22100 Lund, Sweden}

\date{\today{}}

\begin{abstract}
Quantum fluctuation of the energy is studied for an ultracold gas
of interacting fermions trapped in a three-dimensional potential.
Periodic-orbit theory is explored, and energy fluctuations are
studied versus particle number for generic regular and chaotic
systems, as well for a system defined by a harmonic confinement
potential. Temperature effects on the energy fluctuations are
investigated.
\end{abstract}


\maketitle

A trapped gas of ultracold fermionic atoms constitutes an exciting
quantum mechanical many-body system that presently attracts much
attention, see e.g. \cite{Exp}. The confinement potential, as well
as the atomic two-body interaction, can be experimentally
controlled \cite{Grimm}, as can the number of atoms in the trap.
Due to quantum effects the energy is expected to vary, or
fluctuate, in a non-smooth way as the external parameters are
varied. A crucial parameter in the determination of the
fluctuations is the dynamics of the system, that may correspond to
classical regular or chaotic motion. By using semiclassical
methods \cite{Original Article,Mass Article} energy fluctuations
of the many-body system are calculated for generic (quantum)
chaotic and regular systems. We show that a minimum of
fluctuations is obtained by making the dynamics chaotic, while a
harmonic confinement potential generally gives rise to much larger
fluctuations.

The atoms experience an external trapping potential, $V_{trap}$,
and a mutual two-body interaction. For a dilute gas the atom-atom
interaction is well approximated by a zero-range delta potential,
and the many-body Hamiltonian can be written as,
\begin{equation}
 H=\sum_{i=1}^{N}\frac{{\bf p}^2_{i}}{2m}+V_{trap}({\bf r}_{i})+\frac{4\pi\hbar^2a}{m}
 \sum_{i<j}\delta^{(3)}({\bf r}_{i}-{\bf r}_{j}),
\label{Many-body}
\end{equation}
where $a$ is the s-wave scattering length for elastic atom-atom
collisions, that can be experimentally tuned by a magnetic field
through a Feshbach resonance \cite{Timmermans}, from large
negative to large positive values. We restrict the present study
to positive values of $a$, i.e. to repulsive interactions between
the atoms. In addition, we assume an ultracold gas, implying
quantum effects being important. We study a fully unpolarized
system consisting of $N$ fermions with two spin-states, spin-up
($\uparrow$) and spin-down ($\downarrow$). The total density is
then given by, $n\left({\bf r}\right)=n^{\uparrow}\left({\bf
r}\right)+n^{\downarrow}\left({\bf r}\right)=
 2n^{\uparrow}\left({\bf r}\right)$,
where $n^{\uparrow}$ and $n^{\downarrow}$ are the spin-up and
spin-down particle densities. From Eq.(\ref{Many-body}) the
single-particle Hartree-Fock equation is obtained,
\begin{equation}
 \left[-\frac{\hbar^2}{2m}\Delta+gn^{\uparrow}({\bf r})+V_{trap}({\bf r})\right]
 \phi_{i}^{\downarrow}=\epsilon_{i}\phi_{i}^{\downarrow},
\label{Hartree-Fock}
\end{equation}
where we have introduced the interaction parameter
$g=4\pi\hbar^{2}a/m$. The effective mean field potential is thus
given by
\begin{equation}
 V_{{\it {eff}}}=gn^{\uparrow}({\bf r})+V_{trap}({\bf r}).
\label{V_eff}
\end{equation}
The classical dynamics in this effective potential defines the
motion as regular, chaotic or mixed and we expect different
features of the corresponding quantum spectra \cite{Bohigas}. The
diluteness condition of the gas is fulfilled when
$\bar{n}a^{3}\ll1$ where $\bar{n}^{-1/3}$ is the mean
inter-particle spacing.

With given confinement potential and two-body interaction Eq.
(\ref{Hartree-Fock}) can be numerically solved, and the total
energy calculated for a specified number of atoms, see e.g.
Ref.\cite{Supershell}. In this study we are, however, not
interested in a detailed description, but rather in general
features of the system, and the role of the underlying dynamics.
This can be obtained by utilizing semiclassical methods to
calculate the total energy of $N$ confined atoms,
$U(N)=\bar{U}+\tilde{U}$, at low temperature. As is usual in
semiclassics \cite{Brack}, the total energy is divided into a
smoothly varying part $\bar{U}$, and a fluctuating part
$\tilde{U}$, and we shall focus on the non-trivial fluctuating
energy. At low temperatures, $k_{B}T\ll\mu$, where $\mu$ is the
Fermi energy, the fluctuating part can be calculated as
\cite{Original Article},
\begin{equation}
 \tilde{U}(N)=2\hbar^2\sum_{p}\sum_{r=1}^{\infty}\frac{A_{p,r}\kappa_{T}(r\tau_p)}
 {r^2\tau_{p}^2}\cos \left[ rS_{p}/\hbar+\nu_{p,r} \right],
\label{fluctuating energy}
\end{equation}
where the summation $p$ runs over all classical periodic orbits in
the effective mean-field potential, and $r$ is their repetitions.
The amplitude $A_{p,r}$ depends on the stability of the orbit,
$S_{p}\left(E\right)$ is the classical action,
$\tau_{p}=dS_{p}/dE$ the period and $\nu_{p,r}$ the Maslov index.
Temperature effects are included through the function,
\begin{equation}
 \kappa_{T}(\tau)=\frac{\tau/\tau_{T}}{\sinh(\tau/\tau_{T})},
\label{kappa}
\end{equation}
where $\tau_{T}=h/(2\pi^2k_{B}T)$. At zero temperature
$\kappa_T=1$.

The classical functions in Eq.(\ref{fluctuating energy}) are
evaluated at $E=\mu$. Since Eq.(\ref{fluctuating energy}) suffers
from severe convergence problems, we characterize the fluctuating
energy, $\tilde{U}(N)$, by computing its moments. The first
non-trivial moment is the variance, $\sigma^2=\left\langle
\tilde{U}^2\right\rangle$ that gives the typical size of the
fluctuations. This can be written as \cite{Original Article}
\begin{equation}
 \sigma^2=\frac{\hbar^2}
 {2\pi^2}\int_{0}^{\infty}\frac{K(\tau)}{\tau^{4}}\kappa_{T}^2d\tau,
\label{Fluctuating energy form factor}
\end{equation}
where $K(\tau)$ is the diagonal part of the spectral form factor
(the Fourier transform of the two-point energy correlation
function). The form factor is by definition system dependent but
for long times, i.e. $\tau\gg\tau_{min}$ ($\tau_{min}$ is the
period of the shortest periodic orbit), general statistical
properties can be derived \cite{Berry,formfactorchaosI}, namely
\begin{equation}
 K_{reg}=\tau_{H}
\label{K_reg}
\end{equation}
for the regular case, and
\begin{eqnarray}
 K_{ch}=&\left[2\tau-\tau\log\left(1+\frac{2\tau}{\tau_{H}}\right)\right]
 \Theta\left(\tau_{H}-\tau\right)+
\nonumber \\
 &\left[2\tau_{H}-\tau\log\left(\frac{2\tau+\tau_{H}}{2\tau-\tau_{H}}\right)\right]
 \Theta\left(\tau-\tau_{H}\right),
\label{K_chaos}
\end{eqnarray}
for the chaotic case. In these expressions $\tau_{H}=h\bar{\rho}$
is the Heisenberg time and $\Theta$ the Heaviside step function.
The so-called $\tau_{min}$-approximation assumes the smooth
behavior of $K$ (Eqs.(\ref{K_reg}) and (\ref{K_chaos})) all the
way down to $\tau = \tau_{min}$; for $\tau < \tau_{min}$ there are
no periodic orbits and the form factor is zero. In this
approximation simple expressions can be derived for the energy
fluctuations assuming regular or chaotic dynamics, by inserting
the corresponding expressions for the form factor into
Eq.(\ref{Fluctuating energy form factor}). At temperature zero
this gives \cite{Original Article},
\begin{equation}
 \sigma_{reg}^{2}=\frac{b}{24\pi^{4}}E_{c}^{2} \label{sigma_reg}
\end{equation}
for the regular case, where $E_{c}=h/\tau_{min}$ and
$b=E_{c}\bar{\rho}=\tau_H/\tau_{min}$ ("dimensionless
conductance"). In the range $\tau_{min}\ll\tau\ll\tau_{H}$, and
for not too small numbers of particles, it is a good approximation
to write $K_{ch}=2\tau$. In this approximation the energy
fluctuations for chaotic dynamics become,
\begin{equation}
 \sigma_{ch}^{2}=\frac{1}{8\pi^{4}}E_{c}^{2}. \label{sigma_chaos}
\end{equation}
In the studies below we shall, however, generally integrate
Eq.(\ref{Fluctuating energy form factor}) numerically. The
important parameters to determine the fluctuations in energy are
thus the period of the shortest periodic orbit, $\tau_{min}$, and
the mean level spacing at the fermi energy,
$\delta=1/\bar{\rho}=d\mu/dN$. As shortest orbit we assume a
"diameter orbit", i.e. motion along the diameter in the effective
potential, and get $\tau_{min}=4\int_0^{R_{max}}dr/ v(r)$, where
$v(r)$ is the classical velocity and $R_{max}$ corresponding
maximal radius.

At non-zero temperature Eq.(\ref{Fluctuating energy form factor})
is solved numerically using the respective expression for the form
factor, Eqs.(\ref{K_reg}) or (\ref{K_chaos}). In the limit of very
low temperature an analytical expression of the energy
fluctuations may be derived for the chaotic case
($\tau_{T}>>\tau_{H}>>\tau_{min}$) \cite{marc},
\[
 \sigma_{ch}^2
 =\frac{\hbar^2}{2\pi^2}\left[\left(\frac{1}{\tau_{min}}-
 \frac{1}{\tau_{H}}\right)^{2}+\frac{1}{3}\left(\frac{1}{\tau^2_{H}}-
 \frac{1}{\tau^2_{T}}\right)\right],
\]
and similarly for the regular case,
\[
 \sigma_{reg}^2
 =\frac{\hbar^2\tau_{H}}{6\pi^2\tau_{min}}
 \left(\frac{1}{\tau_{min}^2}-\frac{1}{\tau^{2}_{T}}\right).
\]

When the trapping potential is harmonic,
$V_{trap}=m\omega^{2}r^{2}/2$ ($m$ is the atom mass; we shall put
$m$=1), the rather weak two-body interaction between atoms implies
that the effective potential Eq.(\ref{V_eff}) can be approximated
by $\omega_{eff}r^2/2+\epsilon r^4/4$ \cite{Stor Artikel},
implying regular dynamics. Only few periodic orbits appear, and
the expression Eq.(\ref{sigma_reg}), describing a generic regular
system, does not apply. Instead, fluctuations in energy is well
described by two periodic orbits, namely the
circular and diameter periodic orbits \cite{Stor Artikel}. The
interference of the two orbits gives rise to supershell structure,
as noticed in \cite{Supershell}. The semiclassical expression of
the energy in terms of these two periodic orbits reproduces very
well the result obtained from a microscopic Hartree-Fock
calculation.

The second moment of the energy is found by squaring
Eq.(\ref{fluctuating energy}), including only the two shortest
periodic orbits. Only the diagonal terms and those corresponding
to cosines of the action difference contribute, since the others
are eliminated through the averaging procedure due to the rapid
fluctuations in the arguments. The resulting expression for the
energy variance becomes,
\[
\sigma_{2po}^2= \left\langle \tilde{U}^2_{2po}\right\rangle
\approx 2 \hbar^4 \sum_{k=1}^{\infty} \frac{\kappa_T^2}{k^4}
 \left[\frac{A_d^2}{\tau_d^4} + \frac{A_c^2}{\tau_c^4} \right.
\]
\begin{equation}
 \left.-\frac{A_{d}\cdot A_{c}}{\tau_{c}^{2}\tau_{d}^{2}}\left\langle
\cos \left[ k\frac{(S_{d}-S_{c})}{\hbar} \right] \right\rangle
\right], \label{explicit energy}
\end{equation}
where $A_d$ and $A_c$ are amplitudes of the diameter and circle
orbits, respectively \cite{Stor Artikel}. The temperature
dependence simply appears in the pre-factor $\kappa_T$
(Eq.(\ref{kappa})). The first two terms, corresponding to squares
of the two periodic orbits, diverge at the bifurcation point
$\epsilon=0$. Convergence of the expression is, however, restored
by the cross-term.

\begin{figure}[tb]
\includegraphics[width=0.45 \textwidth]{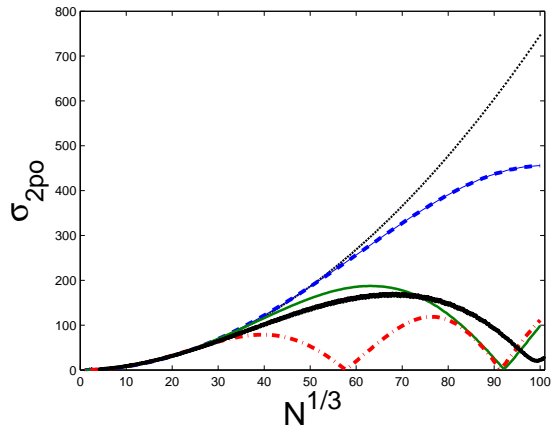}
 \caption{(Color online) Energy fluctuations, $\sigma_{2po}$, versus particle number,
 $N^{1/3}$,
 for the H.O. confinement potential with the two-body interaction strength
 $g=0,\:0.1,\:0.2,\:0.4$ (black dotted, blue dashed, green solid and
 red dashed-dotted). Corresponding Hartree-Fock result for $g$=0.2
 is shown by the thick solid black line.
\label{Explicit Regular for T=3D0}}
\end{figure}

In Fig.\ref{Explicit Regular for T=3D0} we show the energy
fluctuations, $\sigma_{2po}$, as a function of $N^{1/3}$ for some
different interaction strengths, $g=0$ (pure harmonic oscillator;
H.O.), $g=0.1$, $0.2$ and $0.4$. The energy is here (and
throughout the paper) expressed in units of $\hbar \omega$. Energy
fluctuations are seen to decrease with increasing value of $g$.
Supershell structure implies local minima close to
$\sigma_{2po}=0$. For {\it larger} values of $g$ the supernode
appears at {\it smaller} particle numbers. In the case $g=0.4$
also the second supernode is seen in Fig.\ref{Explicit Regular for
T=3D0}. For this interaction strength we also compare to the
Hartree-Fock result, that is seen to give quite similar result.

\begin{figure}[tb]

\begin{minipage}[l]{.22\textwidth}
\hspace{-5mm}
\includegraphics[width=1.1\textwidth]{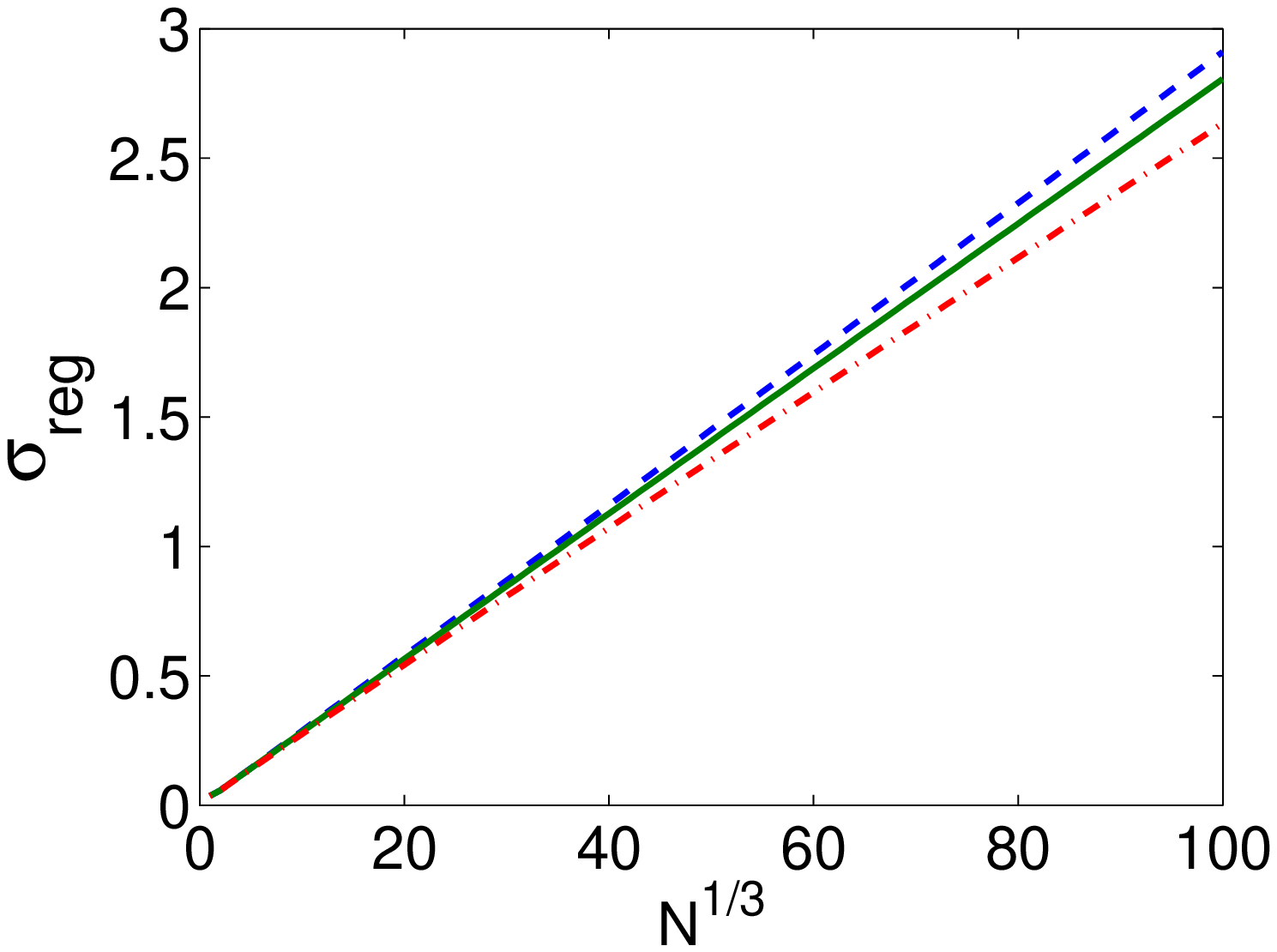}
\end{minipage}
\begin{minipage}[r]{.22\textwidth}
\includegraphics[width=1.1\textwidth]{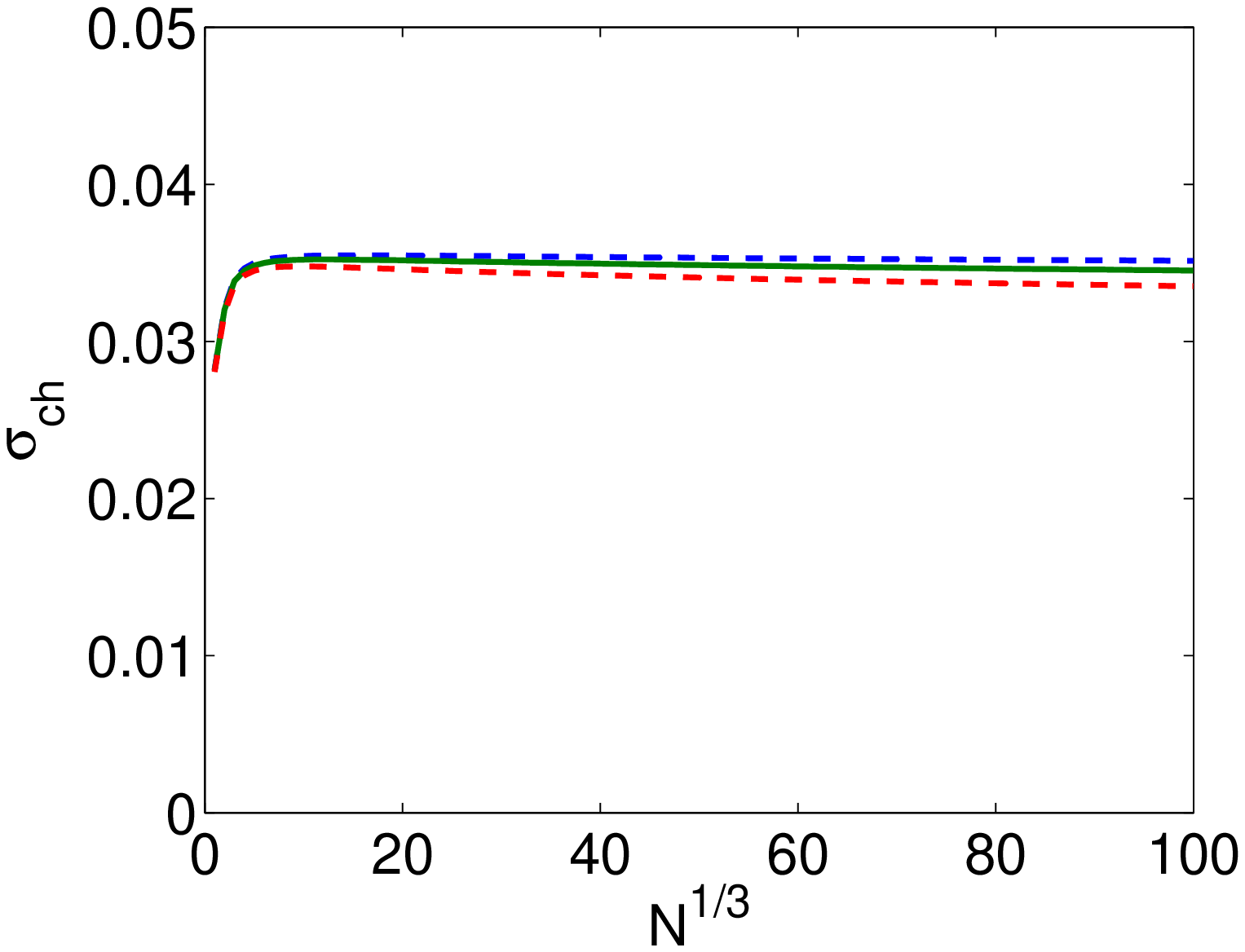}
\end{minipage}
 \caption{(Color online) Energy fluctuations versus particle number, $N^{1/3}$,
 assuming a general regular (left-hand figure) or chaotic (right-hand figure) effective
 potential with the two-body interaction strength  $g=0.1,\:0.2,\:0.4$
 (blue dashed, green solid and red dashed-dotted).
\label{Regular Tmin}}
\end{figure}

Energy fluctuations in a generic regular system is described by
Eq.(\ref{sigma_reg}). Assuming same scaling with particle number
as for the H.O. for $E_C$ and $\bar{\rho}$, gives energy fluctuations as
shown in Fig.\ref{Regular Tmin}. No supershell structure appears,
but $\sigma_{reg}$ is found to increase linearly with $N^{1/3}$.
The effect of the interaction strength is indeed quite small.
Compared to the harmonic confinement potential, Fig.\ref{Explicit
Regular for T=3D0}, the size of the shell energy fluctuations is
notably smaller.

A drastic decrease of energy fluctuations thus appears as the
number of contributing periodic orbits increases from one (family)
orbit for the pure oscillator ($g$=0 case in Fig.\ref{Explicit
Regular for T=3D0}), to two dominant orbits for the harmonic
confinement with an interacting gas ($g > 0$ in Fig.\ref{Explicit
Regular for T=3D0}), and finally to the generic regular system
(Fig.\ref{Regular Tmin}a; e.g. a potential with steep walls ) with
several orbits contributing. The diminishing of energy
fluctuations with increasing number of periodic orbits occurs due
to destructive interferences from several orbits. The special
shell structure with large energy gaps and degeneracies shown by
the harmonic oscillator quantum spectrum, is less pronounced for
systems described by many periodic orbits.

Classical dynamics in the effective potential may become chaotic
by properly arranging the confinement potential. In the purely
chaotic case the number of periodic orbits increases exponentially
with the period time of the orbit and the
$\tau_{min}$-approximation works well. The fluctuations of the
energy originating from chaotic dynamics can thus be calculated by
inserting Eq.(\ref{K_chaos}) into Eq.(\ref{Fluctuating energy form
factor}). In Fig.\ref{Regular Tmin}b energy fluctuations for
chaotic dynamics is shown versus particle number for different
values of the interaction strength. For large particle numbers, $N
> 200$, Eq.(\ref{sigma_chaos}) is a very good approximation of
$\sigma_{ch}$.

In the case of chaotic dynamics, the energy fluctuations are
indeed very small. As compared to the regular case with a harmonic
confinement potential, see Fig. \ref{Explicit Regular for T=3D0},
the fluctuations are about four orders of magnitudes smaller in
the chaotic case. The fluctuations are also seen to be fairly
independent of particle number and on the interaction strength for
$N>100$. The drop in $\sigma_{ch}$ at smaller particle numbers
originates from higher order terms in the expression for $K_{ch}$,
see Eq.(\ref{K_chaos}).

Experimental conditions always implies a non-zero temperature, and
we shall now investigate the temperature dependence of the energy
fluctuations. We assume that the system can be described by the
same density of states at $T>0$ as for $T=0$. This implies that
the fermi energy, the periods of the periodic orbits and the mean
level spacing are all assumed independent of temperature. This
turns out to be a good approximation for temperatures here
studied. The change in temperature thus appears only in the
function $\kappa_T$, Eq.(\ref{kappa}).

The fluctuation in energy at non-zero temperature is obtained by
numerically solving Eq.(\ref{Fluctuating energy form factor}) with
the form factor given by Eq.(\ref{K_reg}) for the regular case,
and by Eq.(\ref{K_chaos}) for the chaotic case. In the case of a
harmonic trapping potential the temperature dependence of the
energy fluctuation is directly given by $\kappa_T$, see
Eq.(\ref{explicit energy}).
\begin{figure}[htb]
\includegraphics[width=0.45\textwidth]{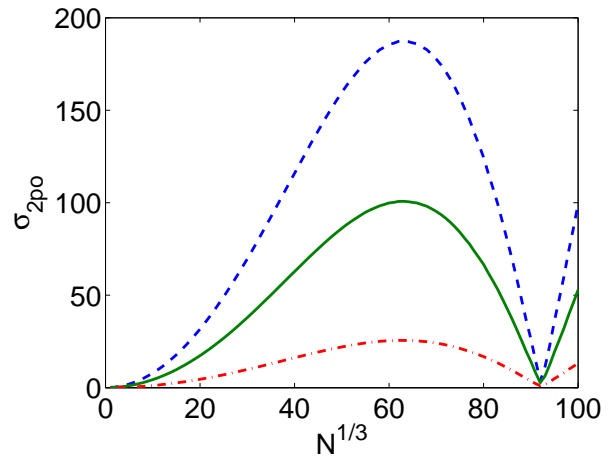}
 \caption{(Color online) Energy fluctuations versus particle number at
 different temperatures, assuming a
 harmonic trapping potential with interaction $g=0.2$.
 The three curves are valid for temperatures
 $k_{B}T/\hbar \omega =0.0,\:0.1,\:0.2$ (blue dashed, green solid and red dashed-dotted).
\label{Regular for T non-zero}}
\end{figure}

In Fig.\ref{Regular for T non-zero} the energy fluctuation is
shown versus particle number for three different temperatures. For
the harmonic trapping potential with $g=0.2$ the supershell
structure is clearly preserved but is suppressed with increasing
temperature.

\begin{figure}[tb]
\includegraphics[width=0.40\textwidth]{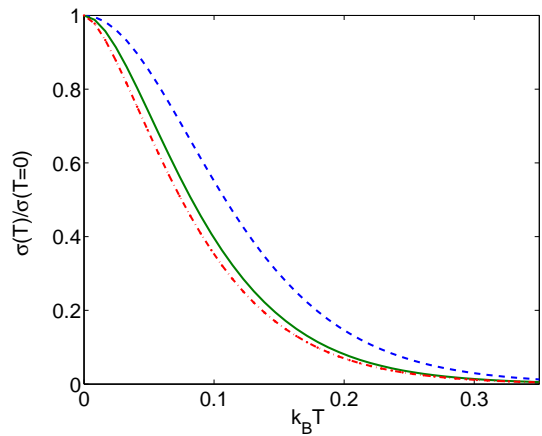}
 \caption{(Color online) Temperature dependence of energy fluctuation relative to zero
 temperature for $g=0.2$. H.O. confinement (blue dashed line), generic
 regular in $\tau_{min}$ (green solid line) and chaotic
 (red dashed-dotted line). Temperature is expressed in units of $\hbar \omega$.}
\label{Max fluct relative T}
\end{figure}

The temperature dependence of the energy fluctuations is shown in
Fig.\ref{Max fluct relative T} for the three studied cases,
regular dynamics in harmonic trap (contribution from two periodic
orbits), generic regular dynamics, and chaotic dynamics. The
fluctuations are shown relative the $T=0$ fluctuations. All three
cases show a similar temperature dependence. At a temperature
$k_BT\approx 0.1 \hbar \omega$ the fluctuations have decreased by
$50\%$, and more or less no energy fluctuations appear for $k_BT >
0.3 \hbar \omega$.

In summary, effects from regular and chaotic dynamics have been
studied for a trapped dilute gas of interacting fermionic atoms.
The fluctuation of the total energy was studied using periodic
orbit theory, and three different dynamical systems were
considered, regular dynamics in a harmonic trap, general regular
dynamics, and chaotic dynamics.

In the case of a harmonic trapping potential a very small number
of periodic orbits are sufficient to describe the fluctuating part
of the energy. If the atoms are not interacting the Hamiltonian
has SU(3) symmetry (pure harmonic oscillator), and one (family)
orbit appears. This gives rise to a monotonically increase of
energy fluctuations with particle number. A week repulsive
interaction between the atoms implies the contribution of two
dominating periodic orbits. The two orbits interfere and give rise
to supershell structure, where the detailed behavior depends on
the interaction strength, $g$. In general, the fluctuations were
found to decrease with interaction strength $g$, and to increase
non-monotonically with particle number.

For a general regular system, as for example an effective
potential with a flat bottom and steep walls, several periodic
orbits contribute. The fluctuations of the energy were then found
to be proportional to $N^{1/3}$, and to be considerably smaller
than for the harmonic trapping potential. The fluctuations are not
sensitive to the interaction strength.

If the effective potential is arranged in such a way that the
dynamics is chaotic, the energy fluctuations are found to be
considerably diminished as compared to the case of regular
dynamics. Compared to the case of a harmonic trap, the
fluctuations for chaotic dynamics were found to be almost four
orders of magnitude smaller for particle numbers $N > 10^4$. This
means that the energy fluctuations more or less vanish for the
chaotic effective potential, independent of interaction strength
and particle number.

Finally, effects of temperature were described in periodic orbit
theory. As expected, a non-zero temperature diminishes the energy
fluctuations. Similar dependence on temperature was found for the
studied different dynamical situations, and the energy fluctuation
was found to be 50\% smaller at temperature $k_B T \approx 0.1$ in
units of $\hbar\omega$, as compared to zero temperature. For a
trap frequency of $10^4 s^{-1}$ temperatures of the order of 10
$nK$ or smaller are thus required to observe significant
fluctuations and supershell structure.

We are grateful to Stephanie Reimann for useful comments.

\end{document}